\documentclass{pasj00}
\begin{document}
\SetRunningHead{J.\ Fukue}
{Radiative Transfer in Accretion Disk Winds}
\Received{yyyy/mm/dd}
\Accepted{yyyy/mm/dd}

\title{Radiative Transfer in Accretion Disk Winds}

\author{Jun \textsc{Fukue}} 
\affil{Astronomical Institute, Osaka Kyoiku University, 
Asahigaoka, Kashiwara, Osaka 582-8582}
\email{fukue@cc.osaka-kyoiku.ac.jp}



\maketitle


\begin{abstract}
Radiative transfer equation in an accretion disk wind 
is examined analytically and numerically
under the plane-parallel approximation
in the subrelativistic regime of $(v/c)^1$,
where $v$ is the wind vertical velocity.
Emergent intensity is analytically obtained
for the case of a large optical depth,
where the flow speed and the source function are
almost constant.
The usual limb-darkening effect,
which depends on the direction cosine 
at the zero-optical depth surface,
does not appear,
since the source function is constant.
Because of the vertical motion of winds, however,
the emergent intensity exhibits
the {\it velocity-dependent} limb-darkening effect,
which comes from the Doppler and aberration effects.
Radiative moments and emergent intensity are also numerically obtained.
When the flow speed is small ($v \leq 0.1c$),
the radiative structure resembles to that of the static atmosphere,
where the source function is proportional to the optical depth,
and the usual limb-darkening effect exists.
When the flow speed becomes large, on the other hand,
the flow speed attains the constant terminal one, and
the velocity-dependent limb-darkening effect appears.
We thus carefully treat and estimate
the wind luminosity and limb-darkening effect,
when we observe an accretion disk wind.
\end{abstract}

\section{Introduction}

Accretion disks are now widely believed
to be energy sources in various active phenomena in the universe:
in protoplanetary nebulae around young stellar objects (YSOs),
in cataclysmic variables (CVs) and supersoft X-ray sources (SSXSs),
in galactic X-ray binaries and microquasars ($\mu$QSOs), and
in active galaxies (ANGs) and quasars (QSOs).
Accretion-disk models have been extensively studies
during these three decades (see Kato et al. 1998 for a review).
Besides the traditional standard disk
by Shakura and Sunayev (1973),
new type disks,
such as advection-dominated accretion flows (ADAF) 
or radiatively-inefficient accretion flows (RIAF)
for the very small mass-accretion rate
(e.g., Narayan, Yi 1994), and
supercritical accretion disks
or so-called slim disks
for the very large mass-accretion rate
(e.g., Abramowicz et al. 1988).

Accretion disk winds have been also extensively examined
in relation to astrophysical jets and outflows:
in bipolar outflows from YSOs,
in mass outflows from CVs and SSXSs,
in relativistic jets from $\mu$QSOs, AGNs, QSOs, and
in gamma-ray bursts (GRBs).
In particular,
intense radiation fields of 
luminous supercritical accretion disks
may be responsible for relativistic jets
 from super-Eddington sources,
such as luminous $\mu$QSOs, GRS~1915$+$105 and SS~433, 
luminous QSOs, 3C~273, and energetic GRBs
(see, e.g., Fukue 2004 for references).

In such circumstances,
radiative transfer in accretion disk winds as well as
accretion disks becomes more and more important.

Radiative transfer in the standard disk has been investigated
in relation to the structure of a static disk atmosphere
and the spectral energy distribution from the disk surface
(e.g., Meyer, Meyer-Hofmeister 1982; Cannizzo, Wheeler 1984).
Furthermore, gray and non-gray models of accretion disks 
were constructed under numerical treatments
(K\v ri\v z and Hubeny 1986; Shaviv and Wehrse 1986;
Adam et al. 1988; Mineshige, Wood 1990; 
Ross et al. 1992; Shimura and Takahara 1993;
Hubeny, Hubeny 1997, 1998; Hubeny et al. 2000, 2001;
Davis et al. 2005; Hui et al. 2005)
and under analytical ones
(Hubeny 1990; Artemova et al. 1996; Fukue, Akizuki 2006a).

Radiative transfer in the accretion disk wind, on the other hand,
has not been well considered
both in the non-relativistic and relativistic regimes.
Recently,
radiative transfer in a moving disk atmosphere
was firstly investigated
in the subrelativistic regime
(Fukue 2005a, 2006a),
and in the relativistic regime
(Fukue 2005b, 2006b; Fukue, Akizuki 2006b).
In contrast to the static atmosphere,
in the moving atmosphere
the boundary condition at the surface of zero optical depth
should be modified (Fukue 2005a, b).
Moreover,
the usual Eddington approximation violates
in the highly relativistic flow (Fukue 2005b;
see also 
Turolla, Nobili 1988; Nobili et al. 1991;
Turolla et al. 1995; Dullemond 1999),
and the velocity-dependent variable Eddington factor
was proposed (Fukue 2006b for a plane-parallel case;
Akizuki, Fukue 2007 for a spherical case).

Radiation hydrodynamical (RHD) simulations were also performed
for radiation-dominated supercritical disks with winds
by several researchers
(Eggum et al. 1985, 1988; Okuda et al. 1997, 2005; 
Okuda, Fujita 2000; Okuda 2002;
Ohsuga et al. 2005; Ohsuga 2006).
In these current studies of RHD simulations for disks and winds,
they were done in the subrelativistic regime
up to the order of $(v/c)^1$, using the moment formalism
and the flux-limited diffusion (FLD) approximation
(Levermore, Pomraning 1981).
The flux-limited diffusion method provides
good approximations to the exact solutions
but only if they are derived from transfer equations
in which terms of the order of $(v/c)^2$ or higher have been retained
(Yin, Miller 1995).

Radiative transfer problems on accretion disks and winds
are not well understood yet,
in particular for the relativistic cases.
Hence, in order clarify the physics,
in addition to RHD simulations,
we must treat the simplified problem
in the analytical way.

In this paper,
we thus examine radiative transfer in the accretion disk wind,
which is assumed to blow off from the luminous disk
in the vertical direction (plane-parallel approximation),
and analytically and numerically obtain 
the flow solutions
for the case without internal heating.
In the previous studies,
radiative flows in the vertically moving atmosphere were solved
in the subrelativistic regime
(Fukue 2005a, 2006a),
and in the relativistic regime
(Fukue 2005b, 2006b; Fukue, Akizuki 2006b),
where only radiative moments were obtained.
In the present paper,
we further obtain the radiation intensity
as well as radiative moments.

In the next section
we describe the basic equations.
In section 3, we show analytical solutions,
while we present numerical solutions in section 4.
The final section is devoted to concluding remarks.


\section{Basic Equations}

Let us suppose a luminous flat disk,
deep inside which
gravitational or nuclear energy is released
via viscous heating or other processes.
The radiation energy is transported in the vertical direction,
and the disk gas, itself, also moves in the vertical direction
as a {\it disk wind}
due to the action of radiation pressure (i.e., plane-parallel approximation).
For simplicity, in the present paper,
the radiation field is considered to be sufficiently intense that
both the gravitational field of, e.g., the central object
and the gas pressure can be ignored.
We also assume the gray approximation,
where the opacities do not depend on the frequency.
As for the order of the flow velocity $v$,
we consider the subrelativistic regime,
where the terms of the first order of $(v/c)$ are retained,
in order to take account of radiation drag.

The radiative transfer equations 
are given in several literatures
(Chandrasekhar 1960; Mihalas 1970; Rybicki, Lightman 1979;
Mihalas, Mihalas 1984; Shu 1991; Kato et al. 1998).
For the plane-parallel geometry in the vertical direction ($z$),
the radiation hydrodynamic equations are described as follows
(Kato et al. 1998).
It should be noted that the basic equations below
are the same as those given in Fukue (2005a),
except for the transfer equation 
for the radiation intensity.

For matter,
the continuity equation is
\begin{equation}
   \rho v = J ~(={\rm const.}),
\label{rho1}
\end{equation}
where $\rho$ is the gas density, $v$ the vertical velocity, and
$J$ the mass-loss rate per unit area.
The equation of motion is
\begin{equation}
   v\frac{dv}{dz} = \frac{\kappa_{\rm abs}+\kappa_{\rm sca}}{c}
                    \left[ F - (E+P)v \right],
\label{v1}
\end{equation}
where $\kappa_{\rm abs}$ and $\kappa_{\rm sca}$
are the absorption and scattering opacities (gray),
which are defined in the comoving (fluid) frame,
$E$ the radiation energy density, $F$ the radiative flux, and
$P$ the radiation pressure in the vertical direction,
which are measured in the fixed (laboratory) frame.
In a gas-pressureless approximation, the energy equation is reduced to
\begin{equation}
   0 = q^+ - \rho \left( j - c\kappa_{\rm abs} E
                  + \kappa_{\rm abs} \frac{2Fv}{c} \right),
\label{j1}
\end{equation}
where $q^+$ is the heating and $j$ is the emissivity,
which is measured in the comoving frame.

For radiation fields,
the frequency-integrated transfer equation,
the zeroth moment equation, and
the first moment equation become, respectively,
\begin{eqnarray}
   \mu \frac{dI}{dz} & = & \rho \left[ 
        \left(1+3\beta\mu \right) \frac{j}{4\pi}
       -\left( \kappa_{\rm abs}+\kappa_{\rm sca} \right)
        \left(1-\beta\mu \right) I 
        \right.
\nonumber \\
        & & \left. + \frac{\kappa_{\rm sca}}{4\pi}
         \{ \left(1+3\beta\mu \right) cE-2F\beta \} \right],
\label{I1.rad}
\\
   \frac{dF}{dz} & = & \rho \left( j - c\kappa_{\rm abs} E
           + \kappa_{\rm abs}F\beta - \kappa_{\rm sca}F\beta \right),
\label{F1.rad}
\\
   \frac{dP}{dz} & = & \frac{\rho}{c}
             \left[ j\beta - (\kappa_{\rm abs}+\kappa_{\rm sca})F
             \right.
\nonumber \\
            && \left.    + \kappa_{\rm abs}cP\beta
                    + \kappa_{\rm sca} \left( cE+cP \right)\beta \right],
\label{P1.rad}
\end{eqnarray}
where $\mu$ is $\cos \theta$,
$\theta$ being the polar angle,
and $\beta = v/c$.
We further adopt the Eddington approximation in the comoving frame,
which is transformed into
\begin{equation}
   P = \frac{1}{3}E + \frac{4}{3}\frac{F}{c}\beta
\label{edd}
\end{equation}
in the fixed frame (Kato et al. 1998).
Here, the transfer equation (\ref{I1.rad}) is corrected
to the order of $(v/c)^1$ (Kato et al. 1998).

Eliminating $j$ with the help of equation (\ref{j1}),
and introducing the optical depth by
\begin{equation}
    d\tau = - ( \kappa_{\rm abs}+\kappa_{\rm sca} ) \rho dz,
\end{equation}
we can rearrange the basic equations up to the order of $(v/c)^1$ as
\begin{eqnarray}
   c^2J\frac{d\beta}{d\tau} &=& -\left( F - 4cP\beta \right),
\label{v}
\\
   \mu \frac{dI}{d\tau} &=& - \frac{1}{4\pi} \frac{ q^+ }
                              { (\kappa_{\rm abs}+\kappa_{\rm sca}) \rho }
                              \left( 1+3\beta\mu \right)
\nonumber \\
                        &&    + \left( 1-\beta\mu \right) I
    - \left( 1+3\beta\mu \right)
                              \frac{1}{4\pi} \left( cE-2F\beta \right),
\label{I}
\\
   \frac{dF}{d\tau} &=& -\frac{ q^+ }
                              { (\kappa_{\rm abs}+\kappa_{\rm sca}) \rho }
                        +F\beta,
\label{F2}
\\
   c\frac{dP}{d\tau} &=& -\frac{ q^+ }
                        { (\kappa_{\rm abs}+\kappa_{\rm sca}) \rho } \beta
                         + F - 4cP\beta,
\label{P2}
\\
 J \frac{dz}{d\tau} &=& -\frac{1}{(\kappa_{\rm abs}+\kappa_{\rm sca}) }c\beta.
\label{z2}
\end{eqnarray}

Finally, 
integrating the sum of equations (\ref{v}) and (\ref{P2})
gives the momentum flux conservation in the present approximation,
\begin{equation}
   c^2J\beta + cP = cP_0 -\int \frac{ q^+ }
              { (\kappa_{\rm abs}+\kappa_{\rm sca}) \rho }\beta d\tau
             = cP_0,
\label{P}
\end{equation}
when there is no heating ($q^+ =0$).
In addition, the subscript 0 means the value
at some reference position (i.e., the wind base).
Similarly from equations (\ref{v}) and (\ref{F2})
we have the energy flux conservation,
\begin{equation}
   \frac{1}{2}Jv^2 + F = F_0 -\int \frac{ q^+ }
              { (\kappa_{\rm abs}+\kappa_{\rm sca}) \rho } d\tau
             = F_0,
\label{F}
\end{equation}
when there is no heating ($q^+ =0$).
Here, the first term on the left-hand side is eventually dropped,
although we retain it here to clarify the physical meanings.

We solve equations (\ref{v}), (\ref{P}), (\ref{F}), and (\ref{I})
for appropriate boundary conditions, and
we obtaine analytic solutions.

As for the boundary conditions at the wind base of $\tau=\tau_0$
and at the wind top of $\tau=0$,
we impose the following conditions.

At the wind base on the disk surface
with an arbitrary optical depth $\tau_0$,
the flow velocity $\beta$ is zero,
the radiative flux is $F_0$
(which is a measure of the strength of radiation field), and
the radiation pressure is $P_0$
(which connects with the radiation pressure gradient),
where the subscript 0 denotes the values at the wind base.

At the wind top, on the other hand,
as already pointed out in Fukue (2005b),
the usual boundary conditions for the static atmosphere
cannot be used for the present radiative wind,
which moves with velocity at the order of the speed of light.
Namely,
the radiation field just above the wind top changes
when the gas itself does move upward,
since the direciton and intensity of radiation
change due to relativistic aberration and Doppler effect
(cf. Kato et al. 1998; Fukue 2000).
If a flat infinite plane with surface intensity $I_{\rm s}$
in the comoving frame is not static,
but moving upward with a speed $v_{\rm s}$ 
($=c\beta_{\rm s}$, and
the corresponding Lorentz factor is $\gamma_{\rm s}$),
where the subscript s denotes the values at the surface,
then, just above the surface,
the radiation energy density $E_{\rm s}$, 
the radiative flux $F_{\rm s}$, and
the radiation pressure $P_{\rm s}$ measured in the inertial frame
become, respectively,
\begin{eqnarray}
   cE_{\rm s} 
   &=& {2\pi I_{\rm s}}
       \frac{3\gamma_{\rm s}^2+3\gamma_{\rm s}u_{\rm s}+u_{\rm s}^2}{3},
\label{Es2}
\\
   F_{\rm s}
   &=& {2\pi I_{\rm s}}
       \frac{3\gamma_{\rm s}^2+8\gamma_{\rm s}u_{\rm s}+3u_{\rm s}^2}{6},
\label{Fs2}
\\
   cP_{\rm s}
   &=& {2\pi I_{\rm s}}
       \frac{\gamma_{\rm s}^2+3\gamma_{\rm s}u_{\rm s}+3u_{\rm s}^2}{3},
\label{Ps2}
\end{eqnarray}
where $u_{\rm s}$ ($=\gamma_{\rm s}v_{\rm s}/c$)
is the flow four velocity at the surface (Fukue 2005b).
As a result, we have the boundary condition
at the wind top within the present approximation,
\begin{equation}
   \frac{cP_{\rm s}}{F_{\rm s}}
      = \frac{2+6\beta_{\rm s}+6\beta_{\rm s}^2}
             {3+8\beta_{\rm s}+3\beta_{\rm s}^2}
      \sim \frac{2}{3} + \frac{2}{9}\beta_{\rm s}.
\label{bc}
\end{equation}
This is the consistent boundary condition
under the present subrelativistic regime
up to the order of $(v/c)^1$,
although Fukue (2005a) have approximately used
the boundary condition,
$cP_{\rm s}/F_{\rm s}=2/3$,
for a static atmosphere.
In general, at the wind top of $\tau=0$,
the boundary condition (\ref{bc}) is not satisfied.
Hence, for given parameters,
we adjust and obtain the mass-loss rate $J$ as an eigen value,
so as to satisfy the boundary condition (\ref{bc}).

As already stated,
for the internal heating in the wind,
we assume that there is no heating source ($q^+=0$),
although it is straightfoward to extend the model
to the case with heating.
On the other hand,
at the wind base on the luminous disk,
there is assumed to be a uniform source of $I_0$.

\section{Analytical Solutions}

In Fukue (2005a),
analytical solutions for radiative moments
were already derived.
For the completeness, in this section,
we first recalculate them under the consistent boundary condition.
Using the analytical expressions for moments,
we then calculate the radiative intensity,
which we wish to know in the present paper.

\subsection{Flow Velocity and Radiative Moments}

Under the present subrelativistic regime,
equation (\ref{F}) means that the radiative flux $F$
is conserved:
\begin{equation}
     F=F_0=F_{\rm s}.
\label{F_sol}
\end{equation}
Using the boundary conditions at the wind base ($v=0$),
equation (\ref{P}) is expressed as
\begin{equation}
     Jv+P=P_0.
\label{P_sol}
\end{equation}

Hence, equation (\ref{v}) becomes
\begin{equation}
   cJ\frac{dv}{d\tau} = -\left( F_{\rm s} - 4P_0 v \right),
\label{1v}
\end{equation}
that can be analytically solved to yield
\begin{equation}
   v = \frac{F_{\rm s}}{4P_0}
     \left[ 1 - e^{\frac{\displaystyle 4P_0}{\displaystyle cJ}
         (\displaystyle \tau - \tau_0)} \right].
\label{v_sol}
\end{equation}
Thus, the radiative flow from the luminous disk without heating
is expressed in terms of the boundary values and the mass-loss rate.
In addition,
the flow velocity $v_{\rm s}$ at the wind top ($\tau=0$) is
\begin{equation}
   v_{\rm s} = \frac{F_{\rm s}}{4P_0}
     \left( 1 - e^{-\frac{\displaystyle 4P_0}{\displaystyle cJ}
         {\displaystyle \tau_0}} \right).
\label{1vs}
\end{equation}

Using the boundary condition at the wind top,
we further impose a condition on the values at boundaries.
Inserting boundary values (\ref{bc}) into momentum equation (\ref{P_sol}),
using equation (\ref{v_sol}),
we have the following relation:
\begin{equation}
    \frac{cP_0}{F_{\rm s}} = \frac{2}{3}
    + \left( \frac{cJ}{4P_0} +\frac{F_{\rm s}}{18cP_0} \right)
     \left( 1 - e^{-\frac{\displaystyle 4P_0}{\displaystyle cJ}
         {\displaystyle \tau_0}} \right).
\label{1bc}
\end{equation}
That is, for given $\tau_0$ and $P_0$ at the wind base,
the mass-loss rate $J$ is determined in units of $F_{\rm s}/c^2$,
as an eigen value.
Compared with Fukue (2005a),
the second term in the first parentheses on the right-hand side
is an additional one,
which appears due to the present corrected boundary condition (\ref{bc}).

As already stated in Fukue (2005a),
the mass-loss rate increases as the initial radiation pressure increases,
while the flow terminal speed increases
as the initial radiation pressure and the loaded mass decrease.

Moreover,
as easily shown from equation (\ref{1bc}),
in order for the flow to exist,
the radiation pressure $P_0$ at the flow base is restricted
in some range,
\begin{equation}
   \frac{2+\sqrt{6}}{6} < \frac{cP_0}{F_{\rm s}} < \frac{2}{3} +  \tau_0,
\end{equation}
which is slightly modified from that given in Fukue (2005a),
due to the corrected boundary condition.
At the upper limit of $cP_0/F_{\rm s} = 2/3 + \tau_0$,
the loaded mass diverges and the flow terminal speed becomes zero.
On the other hand, at the lower limit of $cP_0/F_{\rm s} = (2+\sqrt{6})/6$,
the pressure gradient vanishes, 
the loaded mass becomes zero and the terminal speed approaches
the saturation speed of
\begin{equation}
   \beta_\infty = \frac{-6 + 3 \sqrt{6}}{4} \sim 0.337,
\end{equation}
where the radiative flux $F$ is balanced by the radiation drag $4P_0 v$
with the boundary condition (\ref{bc}).

\begin{figure}
  \begin{center}
  \FigureFile(80mm,80mm){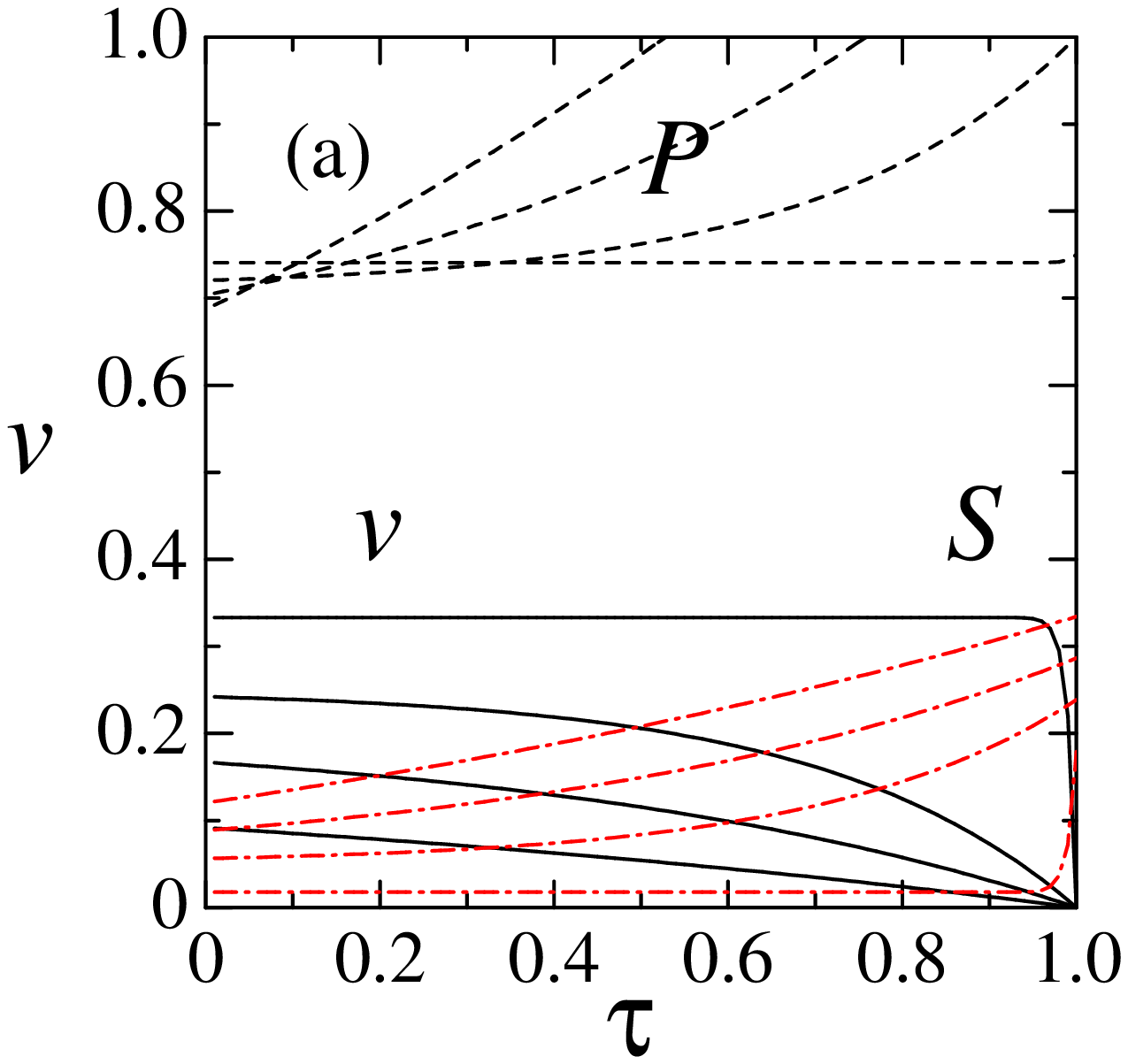}
  \end{center}
  \begin{center}
  \FigureFile(80mm,80mm){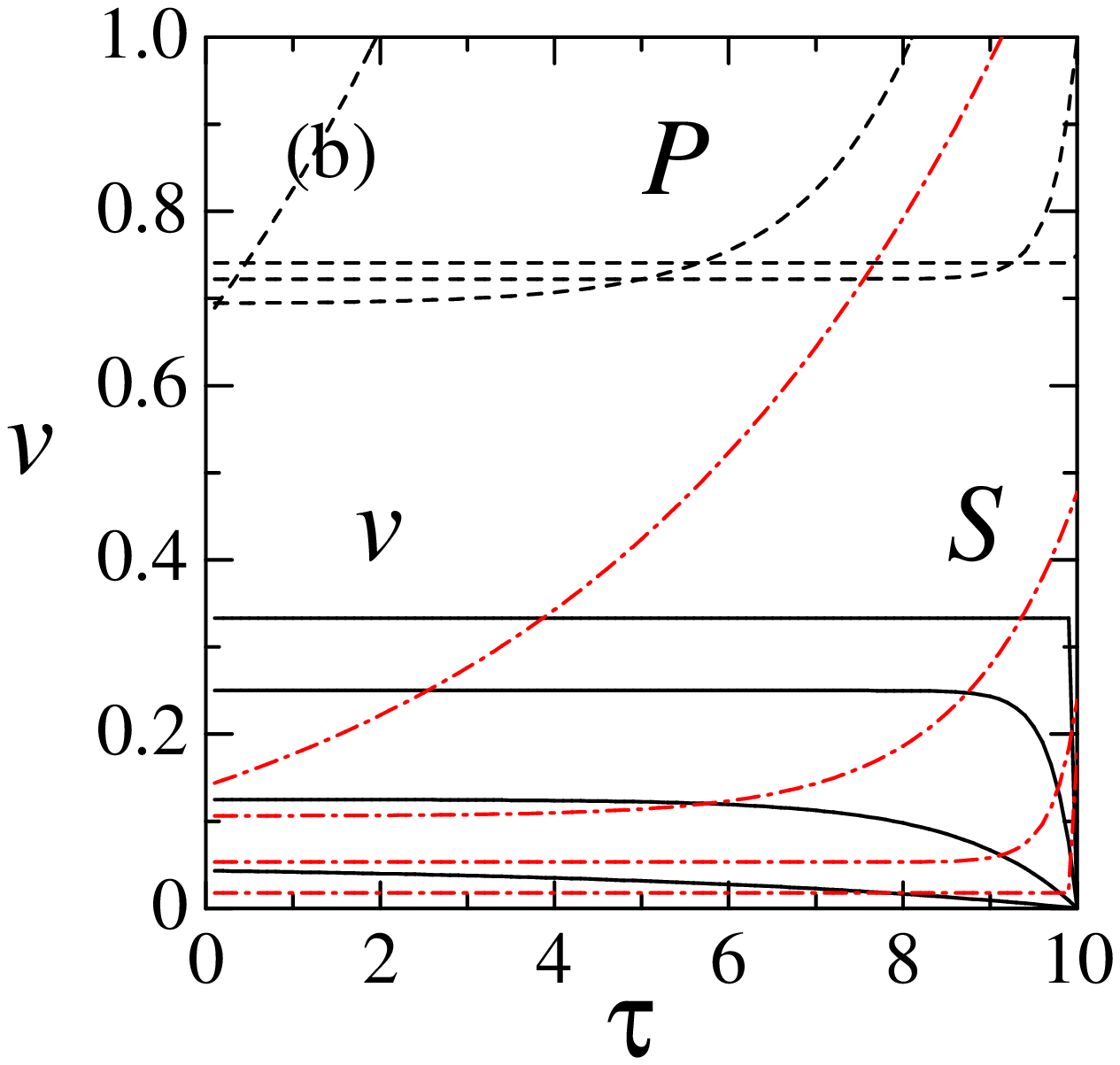}
  \end{center}
\caption{
Flow velocity $v$ (solid curves) in units of $c$,
radiation pressure $P$ (dashed ones) in units of $F_{\rm s}/c$, and
source function $S$ in units of $F_{\rm s}$ (chain-dotted ones)
as a function of the optical depth $\tau$
for several values of $P_0$ at the wind base
in a few cases of $\tau_0$:
(a) $\tau_0=1$ and (b) $\tau_0=10$.
 From top to bottom of $v$ and from bottom to top of $P$ and $S$,
the values of $P_0$ are
0.75, 1, 1.2, 1.4 in (a), and
0.75, 1, 2, 5 in (b).
}
\end{figure}

In figure 1
we show analytical solutions, the flow velocity $v$ (solid curves)
in units of $c$,
the radiation pressure $P$ (dashed ones) in units of $F_{\rm s}/c$, 
and the source function $S$ (chain-dotted ones) in units of $F_{\rm s}$
as a function of the optical depth $\tau$
for several values of $P_0$ at the wind base
in a few cases of $\tau_0$.

When the initial radiation pressure $P_0$ at the wind base is large,
the pressure gradient between the wind base and the wind top is
also large.
As a result, the loaded mass $J$ also becomes large,
but the flow final speed $v_{\rm s}$ is small
due to momentum conservation (\ref{P}).
When the initial radiation pressure $P_0$ is small,
on the other hand,
the pressure gradient becomes small, and
the loaded mass is also small, but
the flow final speed becomes large.
In the latter case,
the source function $S$ becomes almost constant.

Within the present approximation of $(v/c)^1$
with the boundary condition (\ref{bc}),
the flow final speed saturates at $0.337~c$.

\subsection{Emergent Intensity}

Now, we turn to the transfer equation (\ref{I}), or
\begin{equation}
   \mu \frac{dI}{d\tau} = \left( 1-\beta\mu \right) I
    - \left( 1+3\beta\mu \right)
                              \frac{1}{4\pi} \left( cE-2F\beta \right),
\label{II}
\end{equation}
when there is no heating.

Under the present approximation up to the order of $(v/c)^1$,
$F$ is constant, but
$E$ ($P$) and $\beta$ are generally functions of $\tau$.
If, however, the radiation field is sufficiently intense,
the wind flow quickly saturates;
the radiative flux, $F$, is balanced by
the radiation drag force, $(E+P)v$,
and the wind speed reaches the saturation terminal one, $F/(E+P)$.
In such a case,
the wind speed and the radiation quantities are almost constant.
In other words,
the source function, the second term on the right-hand side
of equation (\ref{II}), is almost constant (see figure 1),
since in the present case the source function $S$ is expressed as
\begin{equation}
   S = \frac{cE-2F\beta}{4\pi} = \frac{3cP - 6F\beta}{4\pi}.
\end{equation}

Here, we thus assume that
the wind velocity is constant,
and analytically integrate the transfer equation (\ref{II}).

When the wind velocity becomes a constant terminal one,
using equations (\ref{v_sol}), (\ref{P_sol}), and (\ref{edd}),
we have
\begin{eqnarray}
   \beta &=& \beta_{\rm s} = \frac{F_{\rm s}}{4cP_0},
\\
   cP &=& cP_0 - c^2J\beta
          = \frac{2}{3}F_{\rm s}+\frac{1}{18}\frac{F_{\rm s}^2}{cP_0},
\\
   cE &=& 3cP-4F\beta
          = 2F_{\rm s} - \frac{5}{6}\frac{F_{\rm s}^2}{cP_0}.
\end{eqnarray}
Hence,
\begin{equation}
   cE - 2F\beta 
          = 2F_{\rm s} - \frac{4}{3}\frac{F_{\rm s}^2}{cP_0}
          = 2F_{\rm s} - \frac{16}{3}F_{\rm s}\beta.
\end{equation}
That is, the source function measured in the fixed frame
slightly decreases due to the effect of the relativistic motion.

Under the above situations,
we can now integrate the radiative transfer equation (\ref{II}),
similar to Fukue and Akizuki (2006).
After several partial integrations,
we obtain both an outward intensity $I(\tau, \mu)$ ($\mu>0$)
and an inward intensity $I(\tau, -\mu)$ as
\begin{eqnarray}
   I(\tau, \mu) &=& \frac{cE-2F\beta}{4\pi}\frac{1+3\beta\mu}{1-\beta\mu}
         \left[
                1 - e^{\frac{1-\beta\mu}{\mu}(\tau -\tau_0)}
         \right]
\nonumber \\
     && + I(\tau_0, \mu) e^{\frac{1-\beta\mu}{\mu}(\tau -\tau_0)},
\label{i_sol11} \\
   I(\tau, -\mu) &=& \frac{cE-2F\beta}{4\pi}\frac{1+3\beta\mu}{1-\beta\mu}
         \left[
                1 - e^{-\frac{1-\beta\mu}{\mu}\tau}
         \right],
\label{i_sol12}
\end{eqnarray}
where $I(\tau_0, \mu)$ is the boundary value
at the wind base on the luminous disk.

In general case with finite optical depth $\tau_0$
and uniform incident intensity $I_0$ from the disk,
the boundary value $I(\tau_0, \mu)$ of the outward intensity $I$
consists of two parts:
\begin{equation}
   I(\tau_0, \mu) = I_0 + I(\tau_0, -\mu),
\end{equation}
where $I_0$ ($=F_{\rm s}/\pi$) is the uniform incident intensity and
$I(\tau_0, -\mu)$ is the {\it inward} intensity from
the backside of the disk beyond the midplane.
Determining $I(\tau_0, -\mu)$ from equation (\ref{i_sol12}),
we finally obtain the outward intensity as
\begin{eqnarray}
   I(\tau, \mu) &=& \frac{cE-2F\beta}{4\pi}\frac{1+3\beta\mu}{1-\beta\mu}
         \left[
                1 - e^{\frac{1-\beta\mu}{\mu}(\tau -2\tau_0)}
         \right]
\nonumber \\
     && + I_0 e^{\frac{1-\beta\mu}{\mu}(\tau -\tau_0)},
\nonumber \\
    &\sim& \frac{3F_{\rm s}}{4\pi}
         \left\{ 
           \left( \frac{2}{3}-\frac{16}{9}\beta+\frac{8}{3}\beta\mu \right)
         \left[
                1 - e^{\frac{1-\beta\mu}{\mu}(\tau -2\tau_0)}
         \right]
         \right.
\nonumber \\
   &&     \left.
            + \frac{4}{3} e^{\frac{1-\beta\mu}{\mu}(\tau -\tau_0)}
            \right\},
\label{i_sol13}
\end{eqnarray}
where we have used $F_{\rm s}=\pi I_0$.

Finally, the emergent intensity $I(0, \mu)$ emitted from the wind top
becomes
\begin{eqnarray}
   I(0, \mu) &=& \frac{3F_{\rm s}}{4\pi}
         \left[ 
           \left( \frac{2}{3}-\frac{16}{9}\beta+\frac{8}{3}\beta\mu \right)
         \left(
                1 - e^{-\frac{1-\beta\mu}{\mu}2\tau_0}
         \right)
         \right.
\nonumber \\
   &&     \left.
            + \frac{4}{3} e^{-\frac{1-\beta\mu}{\mu}\tau_0}
            \right],
\nonumber \\
      &\sim& \frac{3F_{\rm s}}{4\pi}
           \left( \frac{2}{3}-\frac{16}{9}\beta+\frac{8}{3}\beta\mu \right)
         ~~~~~{\rm for~large~}\tau_0.
\label{i_sol10}
\end{eqnarray}

Under the present approximation,
where the source function is constant,
the usual limb darkening does not appear:
e.g., $(3F_{\rm s}/4\pi) (2/3+\tau)$ for the Milne-Eddington solution.
Due to the Doppler and aberration effects
originating from the vertical motion of winds, however,
the emergent intensity (\ref{i_sol10})
depends on the wind velocity as well as the direction cosine.
This is the {\it velocity-dependent} limb-darkening effect.

\begin{figure}
  \begin{center}
  \FigureFile(80mm,80mm){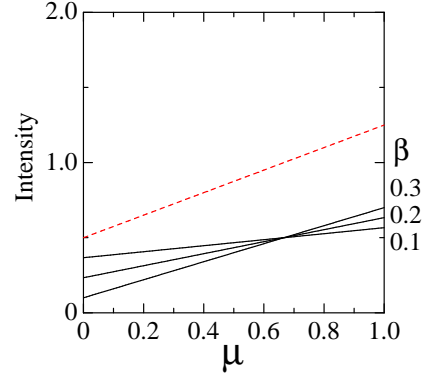}
  \end{center}
\caption{
Normalized emergent intensity
as a function of $\mu$
for the case without heating.
The numbers attached on each curve
are values of $\beta$ of wind velocity.
The dashed straight line is for the usual
Milne-Eddinton solution for the plane-parallel case.
}
\end{figure}

In figure 2,
the emergent intensity $I(0, \mu)$
normalized by the isotropic value $\bar{I}$ ($=F_{\rm s}/\pi$)
is shown for several values of $\beta$
as a function of $\mu$.
Although the present approximation may be valid for $\beta \leq 0.1$,
we show the cases for $\beta \leq 0.3$
in order to stress the velocity-dependency.

As is easily seen in figure 2,
as the velocity becomes large,
the limb-darkening effect becomes prominant.
That is,
the emergent intensity increases in the poleward direction,
while it decreases in the edgeward direction.
As a result,
a wind luminosity would be overestimated
by a pole-on observer and underestimated
by an edge-on observer,
when we observe an optically-thick accretion disk wind.

It is interesting that
the intensity does not change
in the direction of $\mu=2/3$.

\section{Numerical Solutions}

In this section, we numerically solve 
equations (\ref{v}) and (\ref{P2}) [or (\ref{P})]
with constant $F$, and further solve 
equation (\ref{I}) [or (\ref{II})]
for several cases, and
compare the results with that of analytical solutions
to check the accuracy and limitations
of analytical solutions.

\subsection{Flow Velocity and Radiative Moments}

In figure 3
we show several numerical solutions,
the flow velocity $v$ (thick solid curves) in units of $c$
and the radiation pressure $P$ (thick dashed curves) in units of $F_{\rm s}/c$
as a function of the optical depth $\tau$
for several values of $P_0$ at the wind base
in a few cases of $\tau_0$.
Corresponding analytical solutions are shown
by thin curves,
although the mass-loss rates are slightly different
so as to satisfy the upper boundary condition (\ref{bc}).

\begin{figure}
  \begin{center}
  \FigureFile(80mm,80mm){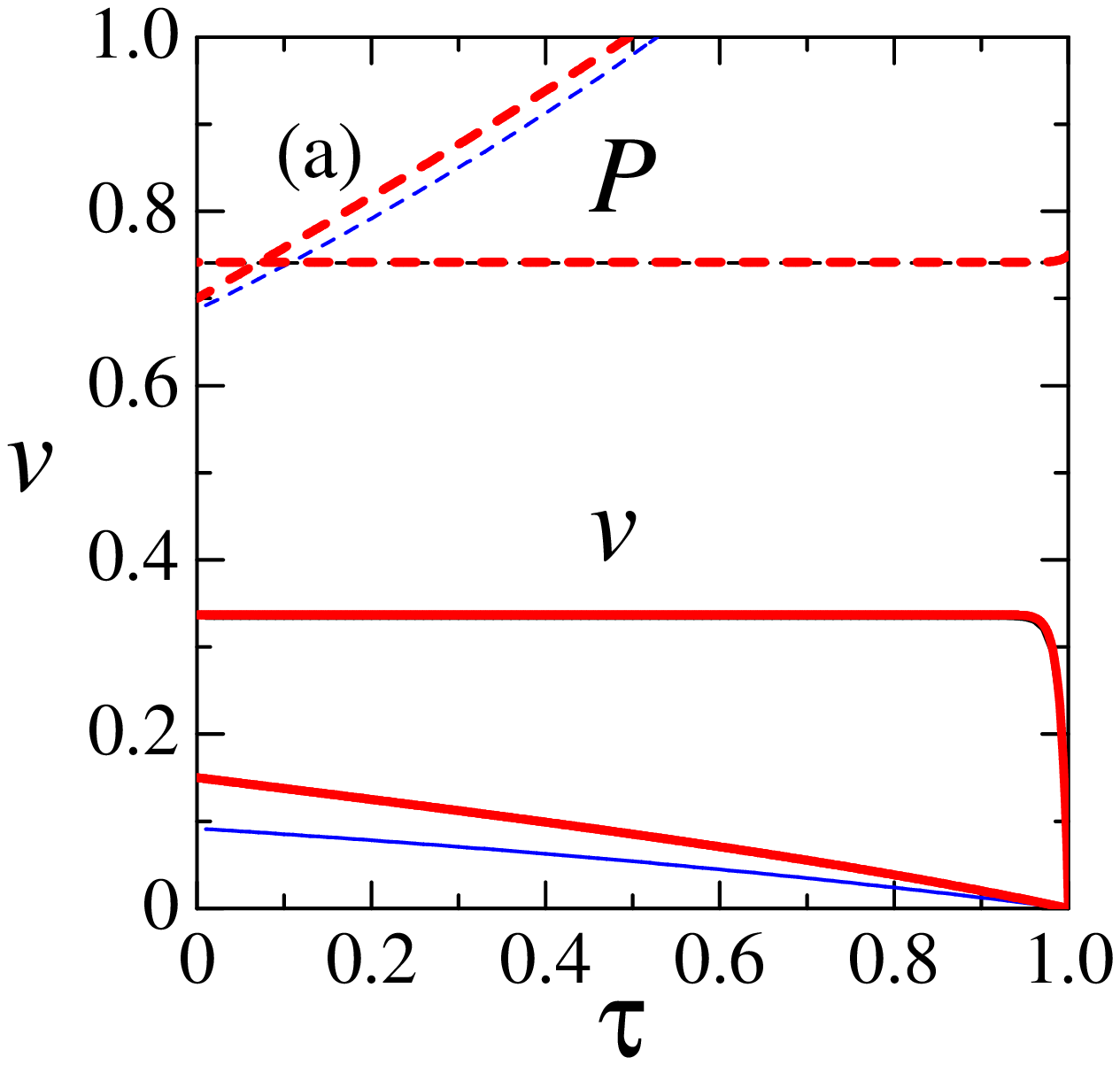}
  \end{center}
  \begin{center}
  \FigureFile(80mm,80mm){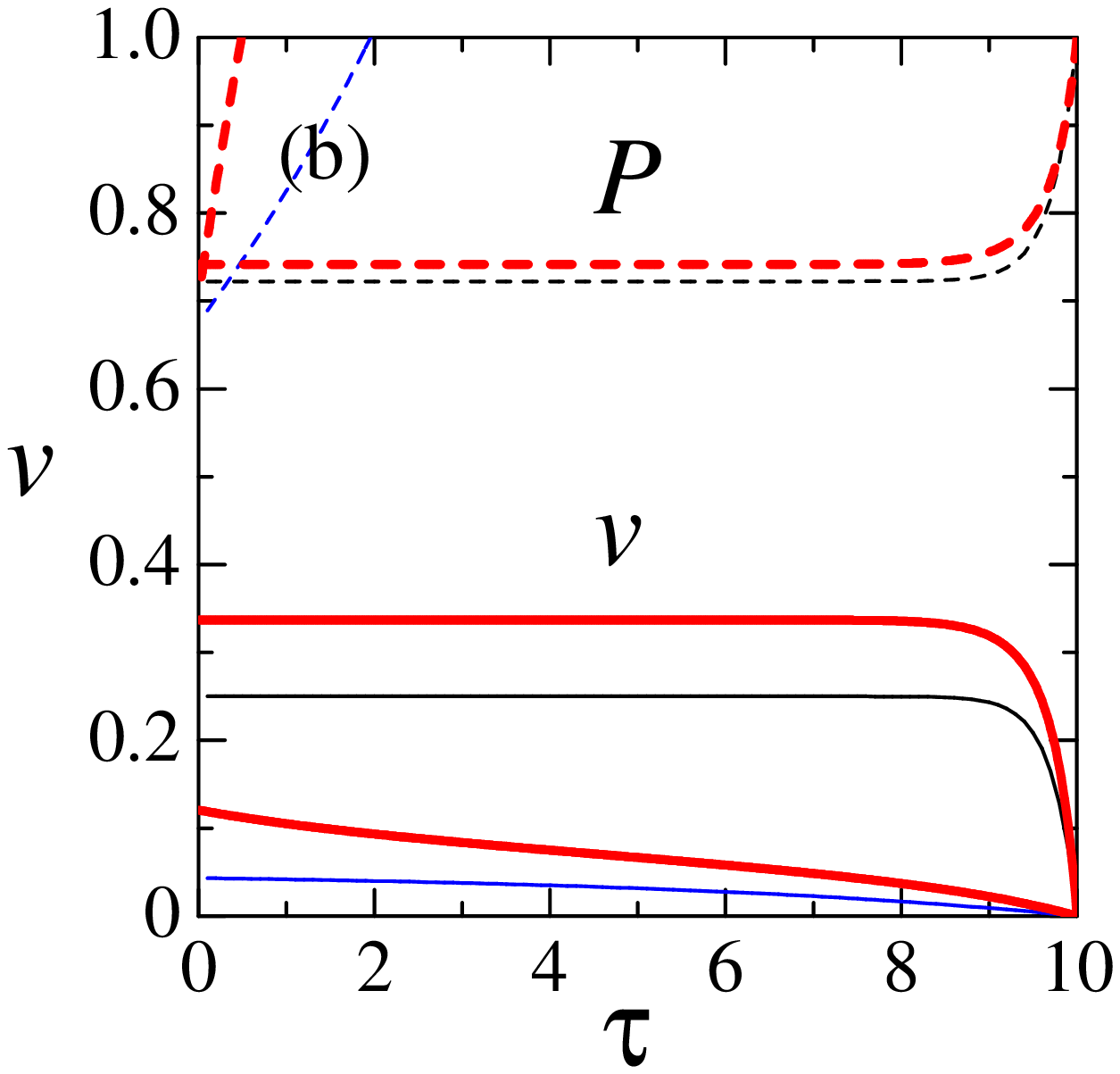}
  \end{center}
\caption{
Flow velocity $v$ (thick solid curves) in units of $c$
and radiation pressure $P$ (thick dashed curves) in units of $F_{\rm s}/c$
as a function of the optical depth $\tau$
for several values of $P_0$ at the wind base
in a few cases of $\tau_0$:
(a) $\tau_0=1$ and (b) $\tau_0=10$.
 From top to bottom of $v$ and from bottom to top of $P$,
the values of $P_0$ are
0.75 and 1.4 in (a), and
1 and 5 in (b).
Corresponding analytical solutions are shown
by thin curves.
}
\end{figure}

As is seen in figure 3, in both analytical and numerical cases,
when the initial pressure $P_0$ and the loaded mass are large,
the flow speed is small ($v \leq 0.1c$).
In such a case,
the radiative structure resembles to that of the static atmosphere,
where the source function is proportional to the optical depth.
When the initial pressure and the loaded mass decrease, on the other hand,
the flow speed becomes large to attain the saturation one.
In such a case,
the source function is almost constant.

In addition, for the same parameters
the analytical and numerical solutions
are slightly different,
although we have solved the same basic equations.
This is understood as follows.
Our basic equations are up to the order of $(v/c)^1$,
and therefore have the accuracy of the same order.
In deriving the analytical solutions,
we have dropped the terms of order of $(v/c)^2$,
while we have retained those terms 
in solving the numerical solutions.
Hence, the analytical and numerical solutions
should be same within the accuracy of $(v/c)^1$.

\subsection{Emergent Intensity}

After obtaining the numerical solutions
for $v$ and $P$ (i.e., $E$),
we can further numerically integrate the transfer equation (\ref{II})
for many angles $\mu$ as a function of $\tau$
under appropriate boundary conditions and
given initial conditions of $\tau_0$ and $P_0$.
As boundary conditions,
we set a uniform source of $I_0$ at the wind base,
while there is no incident intensity at the wind top:
$I^+ (\tau_0, \mu) = I_0 + I^- (\tau_0, -\mu)$ and
$I^- (\tau_0, -\mu) = 0$.
We use meshes of 100 for $\mu$ and 500 for $\tau$.
Finally,
the emergent intensity $I(0, \mu)$ emitted from the wind top
is numerically obtained as a function of angle
for given $\tau_0$ and $P_0$.
Examples of the results are shown in figure 4.

\begin{figure}
  \begin{center}
  \FigureFile(80mm,80mm){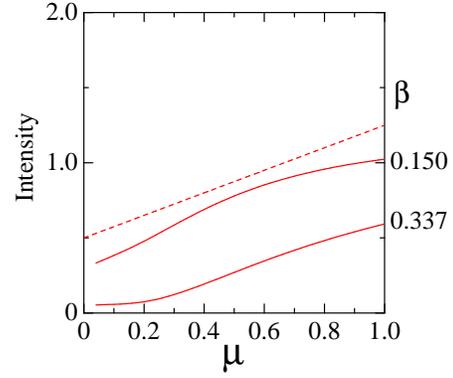}
  \end{center}
\caption{
Normalized emergent intensity
as a function of $\mu$
for the case without heating.
The numbers attached on each curve
are values of $\beta_{\rm s}$ of wind terminal velocity.
The initial optical depth $\tau_0$ is 1 and
the initial pressure $P_0$ is
1.4 ($\beta_{\rm s}=0.150$) and 0.75 ($\beta_{\rm s}=0.337$).
}
\end{figure}

In figure 4,
the emergent intensity obtained numerically is shown 
for the initial optical depth of $\tau_0=1$,
and for several values of $\beta_{\rm s}$,
that is attached on each curve.
When the terminal speed is small, as already stated,
the wind structure resembles to that of the static atmosphere.
In the present case, however, the optical depth of wind is finite.
As a result, at around $\mu \sim 1$
the emergent intensity is almost constant
with the values of $I_0$ (i.e., the peaking effect diminishes),
while the usual limb-darkening effect recovers for small $\mu$,
where the line-of-sight length is long.

When the terminal speed is large, on the other hand,
the velocity-dependent limb-darkening effect appears
again for the numerical solutions.
That is to say, as stated in figure 2,
due to the Doppler and aberration effects
originating from the vertical motion of winds, 
the emergent intensity 
depends on the wind velocity as well as the direction cosine.
In particular,
the emergent intensity is enhanced and increases
as $\mu$ increases.

By the way,
in figure 4
the emergent intensity entirely decreases
for all $\mu$ as the velocity increases.
Apparently, this seems to be curious,
because the relativistic boosts work
in the regime with large velocity.
The reason is also the relativistic effect.
The present subrelativistic flow
up to the order of $(v/c)^1$,
the momentum and energy of radiation fields
accelerate the flow.
Although the flux $F$ is conserved
within the present order of $(v/c)^1$,
the radiation intensity diminishes
due to the interaction between the field and the flow
[the $2F\beta$ term in the second term
on the right-hand side of equation (\ref{II})].
In other words,
the source term effectively decreases.
As a result,
the emergent intensity entirely decreases,
compared with the non-relativistic case.

\begin{figure}
  \begin{center}
  \FigureFile(80mm,80mm){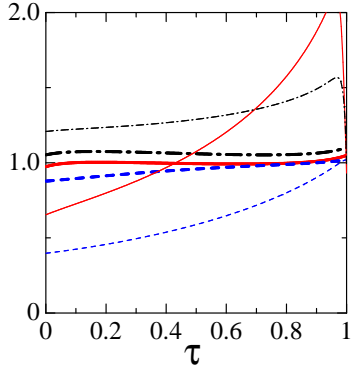}
  \end{center}
\caption{
Ratios between the quantities obtained from the moment equations
and those obtained from the intensity.
Solid curves and dashed ones denote
the energy density and the flux, respectively.
In the small velocity case of $\beta_{\rm s}=0.150$ (thick curves),
both quantities are almost consistent.
In the large velocity case of $\beta_{\rm s}=0.337$ (thin curves),
on the other hand, the consistency becomes worse.
In addition,
chain-dotted curves mean the three times Eddington factor
for the quantities obtained from the intensity.
}
\end{figure}

We here briefly check the consistency of the numerical results.
In order to obtain the radiation intensity $I$ 
from equation (\ref{II}),
we use the radiation energy density $E_{\rm mom}$
and the radiative flux $F_{\rm mom}$ ($=F_{\rm s}$)
obtained from the moment equations (\ref{P}).
Once the radiation intensity $I(\tau, \mu)$ is obtained,
we can calculate the radiation energy density $E_{\rm sim}$
and the radiative flux $F_{\rm sim}$ by the definition.
If the quantities obtained from the moment equations
and those obtained from the intensity coincide each other,
the solutions are consistent, and vice versa.

In figure 5
we plot the ratios
$E_{\rm sim}/E_{\rm mom}$ (solid curves) and
$F_{\rm sim}/F_{\rm mom}$ (dashed ones)
for the small velocity case of $\beta_{\rm s}=0.150$ (thick curves)
and
for the large velocity case of $\beta_{\rm s}=0.337$ (thin curves).
In addition, we also plot the quantities
$(E_{\rm sim}+4F_{\rm sim}v)/(3P_{\rm sim})$,
the three times Eddington factor,
by chain-dotted curves.

As is seen from figure 5,
in the case of small velocity,
where the structure resembles to that of the static case,
the ratios are almost unity
and the solutions are consistent.
In the small velocity case, furthermore,
the Eddington factor is almost $1/3$.
In the case of large velocity, on the other hand,
the ratios differ from unity.
The main reason of this discrepancy is the relativistic effect.
As already stated,
the radiation intensity diminishes
due to the interaction between the field and the flow.
As a result,
the intensity $I(\tau, \mu)$ as well as 
the emergent intensity $I(0, \mu)$ entirely decreases.
Thus, $F_{\rm sim}$ becomes smaller than $F_{\rm mom}$.
Although the reason that the energy density ratio increases
is less clear,
it seems to be the relativistic peaking effect
(Doppler boost and relativistic abberation).
Indeed, the Eddington factor in the large velocity case
is somewhat larger than unity.
This would be also the relativistic peaking effect.
Anyway, as the velocity becomes large,
the consistency becomes worse,
and the present treatment under the order of $(v/c)^1$
would be invalid.


\section{Concluding Remarks}

In this paper 
we have examined the radiative transfer problem
in an accretion disk wind
under the plane-parallel approximation
in the subrelativistic regime of $(v/c)^1$.
The flow velocity, the radiation pressure distribution, 
and other quantities
are analytically and numerically
solved as a function of the optical depth
for the case without heating.
Furthermore,
the emergent intensity from the wind top with zero optical depth
is also analytically obtained
under the assumption of constant flow velocity,
while numerically calculated.

When the source function is constant,
te usual limb-darkening effect does not appear.
However,
the emergent intensity exhibits
the {\it velocity-dependent} limb-darkening effect,
which originates from the Doppler and aberration effects,
associating with the vertical subrelativistic motion of winds.
As a result,
a wind luminosity would be overestimated by a pole-on observer
and underestimated by an edge-on observer,
when we observe an optically thick accretion disk wind.

It should be noted that
the {\it apparent} optical depth
in the relativistically moving media.
Abramowicz et al. (1991) pointed out that
the optical depth in the relativistic flow
decreases as $\gamma(1-\beta\mu)\tau$ toward the downstream direction,
due to the Doppler and aberration effects.
Inspecting equation (\ref{II}) or solution (\ref{i_sol10}),
we find that,
in the present subrelativistic flow,
the optical depth $\tau$ is apparently replaced by
$(1-\beta\mu)\tau$.
This is just consistent with the results
by Abramowicz et al. (1991)
within the order of $(v/c)^1$.

In the present modelized plane-parallel geometry,
the gas density at the wind top is finite,
since the wind velocity there is finite.
Rigorously speaking, this situation is inconsistent
with the definition of the wind top,
where the optical depth is effectively zero.
In more realistic cases,
the disk wind would geometrically diverge,
and the density quickly drops.
As a result, the surface of a moving photosphere exists,
even if the wind terminal velocity is finite.
The present model mimics such a realistic disk wind,
where the velocity-dependent limb darkening effect
would also appear.
In other words, as long as
the acceleration of wind takes place in a small height,
and then the wind diverges,
the present assumption would be valid.

The radiative transfer problem investigated in the present paper
must be quite {\it fundamental problems} for
accretion disk physics and astrophysical jet formation.
In this paper, for simplicity,
we only considered the subrelativistic case of $(v/c)^1$.
The fully relativistic case of $(v/c)^2$
will be explored in the future.

\vspace*{1pc}

The author would like to thank an anonymous referee for valuable comments.
This work has been supported in part
by a Grant-in-Aid for Scientific Research (18540240 J.F.) 
of the Ministry of Education, Culture, Sports, Science and Technology.


\end{document}